\documentclass[runningheads,a4paper]{llncs} 

\usepackage{times}
\usepackage{helvet}
\usepackage{courier}

\usepackage{flafter}
\usepackage{graphicx}
\graphicspath{{figs/}}
\usepackage{colortbl} 
\definecolor{grey}{rgb}{0.95,0.95,0.95}
\definecolor{darkgrey}{rgb}{0.6,0.6,0.6}

\usepackage[T1]{fontenc}
\usepackage[utf8]{inputenc}
\fontfamily{ptm}\selectfont
\usepackage[iso]{umlaute}

\usepackage{longtable}
\usepackage{float}
\usepackage{amsmath}
\usepackage{amsfonts}
\usepackage{amssymb}
\usepackage{amscd}
\usepackage{bbm}

\usepackage{algorithm}
\usepackage{algorithmic}
\usepackage{mathrsfs}
\usepackage{wrapfig}
\usepackage{paralist}
\usepackage{url}
\usepackage{xspace}
\usepackage{rotating}
\usepackage{multirow}

\usepackage{subfig}
\usepackage[utf8]{inputenc}
\usepackage{xspace}
\usepackage{bbm}
\usepackage{url}

\newcommand{\eg}{e.\,g.,\xspace}
\newcommand{\ie}{i.~e.,\xspace}

\newcommand{\flickr}{flickr\xspace}
\newcommand{\bibs}{BibSonomy\xspace}
\newcommand{\twitter}{Twitter\xspace}
\newcommand{\facebook}{facebook\xspace}
\newcommand{\nameling}{Nameling\xspace}
\newcommand{\wikipedia}{Wikipedia\xspace}
\newcommand{\wiktionary}{Wiktionary\xspace}

\newcommand{\keywords}[1]{\par\addvspace\baselineskip
\noindent\keywordname\enspace\ignorespaces#1}

\newcommand{\comments}[1]{}
\newcommand{\todo}[1]{}

\newcommand{\qap}{QAP\xspace}

\newcommand{\cov}{\ensuremath{\mathop{cov}}}
\newcommand{\var}{\ensuremath{\mathop{var}}}

\newcommand{\CN}{\ensuremath{\mathop{C\!N}}\xspace}
\newcommand{\wCN}{\ensuremath{\widetilde{\mathop{C\!N}}}\xspace}
\newcommand{\JC}{\ensuremath{\mathop{JC}}\xspace}
\newcommand{\wJC}{\ensuremath{\widetilde{\mathop{JC}}}\xspace}
\newcommand{\PA}{\ensuremath{\mathop{P\!A}}\xspace}
\newcommand{\wPA}{\ensuremath{\widetilde{\mathop{P\!A}}}\xspace}
\newcommand{\AC}{\ensuremath{\mathop{A\!A}}\xspace}
\newcommand{\wAC}{\ensuremath{\widetilde{\mathop{A\!A}}}\xspace}
\newcommand{\RA}{\ensuremath{\mathop{R\!A}}\xspace}
\newcommand{\wRA}{\ensuremath{\widetilde{\mathop{R\!A}}}\xspace}
\newcommand{\NM}{\ensuremath{\mathop{N\!EW}}\xspace}
\newcommand{\CS}{\ensuremath{\mathop{C\!O\!S}}\xspace}
\newcommand{\wCS}{\ensuremath{\widetilde{\mathop{C\!O\!S}}}\xspace}
\newcommand{\PR}{\ensuremath{\mathop{P\!R}}\xspace}

\newcommand{\shuffle}[1]{\ensuremath{\underline{#1}}}
\newcommand{\rewire}[1]{\ensuremath{\overline{#1}}}

\newcommand{\coloneqq}{\mathrel{\mathop{:}}=}       
\newcommand{\walk}[3]{\ensuremath{#1\rightarrow_{#2}#3}}

\newcommand{\R}{\mathbb{R}}

\newcommand{\size}[1]{\ensuremath{|#1|}}

\newcommand{\set}[1]{\ensuremath{\{#1\}}}

\newcommand{\Nen}{\ensuremath{G^N_{\text{\textsc{EN}}}}\xspace}
\newcommand{\Nde}{\ensuremath{G^N_{\text{\textsc{DE}}}}\xspace}
\newcommand{\Nfr}{\ensuremath{G^N_{\text{\textsc{FR}}}}\xspace}

\newcommand{\Cen}{\ensuremath{G^C_{\text{\textsc{EN}}}}\xspace}
\newcommand{\Cde}{\ensuremath{G^C_{\text{\textsc{DE}}}}\xspace}
\newcommand{\Cfr}{\ensuremath{G^C_{\text{\textsc{FR}}}}\xspace}

\newcommand{\CTweets}{\ensuremath{G^C_{\text{Twitter}}}\xspace}
\newcommand{\NTweets}{\ensuremath{G^N_{\text{Twitter}}}\xspace}

\newcommand{\Group}{Group}
\newcommand{\Friend}{Friend}

\newcommand{\cooc}{co-occurrence}

\begin{document}
\setlength{\abovecaptionskip}{0pt}
\setlength{\abovedisplayskip}{1pt}
\setlength{\belowdisplayskip}{1pt}

\mainmatter  

\title{Onomastics 2.0 \\[0.5\baselineskip]
\large{The Power of Social Co-Occurrences}}
\titlerunning{Onomastics 2.0}

\author{Folke Mitzlaff\inst{1} \and Gerd Stumme\inst{2}}
\authorrunning{Mitzlaff et al.} 

\tocauthor{Folke Mitzlaff and Gerd Stumme}
\institute{
        \email{mitzlaff@cs.uni-kassel.de}
        \and
        \email{stumme@cs.uni-kassel.de}\\
Knowledge and Data Engineering Group (KDE), University of Kassel\\
Wilhelmsh\"oher Allee 73, D-34121 Kassel, Germany
}

\comments{
\toctitle{Lecture Notes in Computer Science}
\tocauthor{Authors' Instructions}
}
\maketitle

\begin{abstract}
  Onomastics is ``\emph{the science or study of the origin and forms
    of proper names of persons or places.}''\footnote{``Onomastics.''
    \emph{Merriam-Webster.com}. 2013. \url{http://www.merriam-webster.com}
    (11 February 2013).} Especially personal names play an important
  role in daily life, as all over the world future parents are facing
  the task of finding a suitable given name for their child.  This
  choice is influenced by different factors, such as the social
  context, language, cultural background and, in particular, personal
  taste. \comments{Concomitant with the globalization, the importance
    of traditional naming habits decreases and new naming patterns
    emerge.}

  With the rise of the Social Web and its applications, users more and
  more interact digitally and participate in the creation of
  heterogeneous, distributed, collaborative data collections. These
  sources of data also reflect current and new naming trends as well
  as new emerging interrelations among names.

  The present work shows, how basic approaches from the field of
  social network analysis and information retrieval can be applied for
  discovering relations among names, thus extending Onomastics by data
  mining techniques. The considered approach starts with building
  co-occurrence graphs relative to data from the Social Web,
  respectively for given names and city names. As a main result,
  correlations between semantically grounded similarities among names
  (\eg geographical distance for city names) and structural graph
  based similarities are observed.

  The discovered relations among given names are the foundation of the
  \nameling\footnote{\url{http://nameling.net}}, a search engine and
  academic research platform for given names which attracted more than
  30,000 users within four months, underpinning the relevance of the
  proposed methodology.

\end{abstract}

\enlargethispage*{3\baselineskip}

\section{Introduction}\label{sec:intro}
Most future parents face the challenge of finding a suitable given
name for their child. Many non-technical influence factors have to be
considered, such as cultural background, social environment, personal
preference and current trends. Some factors may even be contradictory,
\eg considering the personal preference of both parents. Even if both
parents agree on a favorite given name, often the social environment
prevents a final decision, if many children in the neighborhood are
given the preferred name. Typically, parents end up browsing through
endless lists of thousands of given names, although only a small
fraction of those names are ``relevant'', considering, \eg the
cultural background and personal preference.

From a technical point of view, the scenario described above forms a
challenging recommender setting, where for a given social context (\eg
the parents' given names, hometown, friends), a list of relevant names
is requested. With the rise of the so called ``social web'', many
sources for background information became available, covering social
interaction (\eg \facebook\footnote{\url{http://www.facebook.com}}),
encyclopedic knowledge (\eg
\wikipedia\footnote{\url{http://www.wikipedia.org}}) and short
personal messages (\eg \twitter\footnote{\url{http://twitter.com}}).

The present work tackles the task of recommending given names based on
data from the social web by analyzing relations among names which are
derived from word co-occurrences in \wikipedia and \twitter. Different
well known basic approaches for determining the similarity of words
are applied and evaluated. The obtained results already gave raise to
the \nameling, a search engine for given names which attracted more
than 30,000 users within less than four months, underpinning the
practical relevance of the discovered relations.

The experiments on name relatedness are preceded by an in-depth
comparative analysis of the underlying co-occurrences networks, giving
insights into the interrelation of networks derived from different
language editions of \wikipedia. The proposed methodological approach
can also be seen as a general set up for analyzing and evaluating
co-occurrence networks of named entities and respective similarity
metrics. Exemplarily, all experiments are conducted in parallel on
city names.

This work is structured as follows: Section~\ref{sec:related} gives an
overview on related topics and respective works.
Section~\ref{sec:preliminaries} summarizes relevant basic concepts and
notations, Section~\ref{sec:data} and~\ref{sec:networks} describe the
underlying co-occurrence networks and their data sources, together
with a comparative analysis of the networks. In
Section~\ref{sec:task}, various similarity functions are described and
evaluated, Section~\ref{sec:closing} finally summarizes the obtained
results and points towards future work.
\enlargethispage*{3\baselineskip}
\section{Related Work}\label{sec:related}
\todo{
  * KEYWORDS: distributional semantics, corpora analysis, semantic relatedness

  * named entity disambiguation
  * person name vs. given name
  * 
  * ESA, WikiRelate

  * geographic co-occurrences

  CONTRIBUTION
  -> focus on evaluation of performance of building blocks
  ->

  ``Using Encyclopedic Knowledge for Named Entity Disambiguation''
  -> detect entities -> disambiguation using
     word contexts -> similarities

  ``WikiRelate''

  ``ESA''

  ``Large-Scale Named Entity Disambiguation Based on Wikipedia Data''

==== Statistical semantics
 - Second Order Co-occurrence PMI for Determining the Semantic
 Similarity of Words

 - 

==== Semantic similarity or semantic relatedness

==== Link Prediction / Vertex Similarity

}%
Early applications of data mining techniques for the analysis of place
names include~\cite{leino2003discovery}, where spatial data of lakes
in Finland is analyzed. The application of personal names for
estimating ethnicity for census data using data mining is presented
in~\cite{mateos2007ontology}. The task of identifying different
variants of named entities is extensively studied, examples
include~\cite{bikel1999algorithm,jiang2007named,steinberger2007crosslingual,jijkoun2008named,hermjakob2008translation,nothman2013learning}

The present work aims at discovering and assessing new emergent
relations among given names based on data from the social
web. Methodologically, the considered approach is closely related to
work on \emph{distributional similarity} where, more generally,
semantic relations among named entities are investigated. However,
this work presents an approach to the discovery and analysis of
relatedness from a social network analyst's point of view, which is
connected to the field of \emph{link prediction} and (more generally)
\emph{vertex similarity} in graphs. The proposed methodology is
complementary applied for analyzing the relatedness of city names
which relates to work published on Geographic Information
Retrieval~\cite{overell2007geographic}.

\paragraph{Distributional Similarity \& Semantic Relatedness:}
The field of distributional similarity and semantic relatedness has
attracted a lot of attention in literature during the past decades
(see \cite{cohen2009empirical} for a review). Several statistical
measures for assessing the similarity of words are proposed, as for
example in~\cite{lesk1969word,grefenstette1992finding,islam2006second,landauer1997solution,turney2001mining}. Notably,
first approaches for using \wikipedia as a source for discovering
relatedness of concepts can be found
in~\cite{bunescu2006using,strube2006wikirelate,gabrilovich2007computing}.

\paragraph{Vertex Similarity \& Link Prediction:}
In the context of social networks, the task of predicting (future)
links is especially relevant for online social networks, where social
interaction is significantly stimulated by suggesting people as
contacts which the user might know. From a methodological point of
view, most approaches build on different similarity metrics on pairs
of nodes within weighted or unweighted
graphs~\cite{jeh2002simrank,leicht2005vertex,lu2009similarity,lu2010prediction}. A
good comparative evaluation of different similarity metrics is
presented in \cite{libennowell2007linkprediction}.

The present work combines approaches from both the link prediction and
semantic relatedness tasks with a focus on the structural analysis of
the underlying co-occurrence networks and their inter network
correlations. Relatedness is considered only for a single class of
entities, respectively given names and city names and the obtained
results are evaluated in a novel experimental setup which gives also
insights into the underlying network structure.

\section{Preliminaries}\label{sec:preliminaries}
In this chapter, we want to familiarize the reader with the basic
concepts and notations used throughout this paper.

A \emph{graph} $G=(V,E)$ is an ordered pair, consisting of a finite
set $V$ of \emph{vertices} or \emph{nodes}, and a
set $E$ of \emph{edges}, which are two-element subsets of $V$. A
\emph{directed graph} is defined accordingly: $E$ denotes a subset of
$V\times V$. For simplicity, we write $(u,v)\in E$ in both cases for
an edge belonging to $E$ and freely use the term \emph{network} as a
synonym for a graph. In a \emph{weighted Graph} each edge $l\in E$ is
given an edge weight $w(l)$ by some weighting function $w\colon
E\rightarrow\R$. For a subset $U\subseteq V$ we write ${G}_{|U}$ to
denote the sub graph \emph{induced by $U$}. The \emph{density} of a
graph denotes the fraction of realized links, \ie $\frac{2m}{n(n-1)}$
for undirected graphs and $\frac{m}{n(n-1)}$ for directed graphs
(excluding self loops).
The \emph{neighborhood} $\Gamma$ of a node $u\in V$ is the set of
\emph{adjacent} nodes $\set{v\in V\mid (u,v)\in E}$. The \emph{degree}
of a node in a network measures the number of connections it has to
other nodes. For the \emph{adjacency matrix} $A\in\R^{n\times n}$ with
$n = |V|$ holds $A_{ij}=1$ ($A_{ij}=w(i,j)$) iff $(i,j)\in E$ for any
nodes $i,j$ in $V$ (assuming some bijective mapping from
${1,\ldots,n}$ to $V$). We represent a graph by its according
adjacency matrix where appropriate.

A \emph{path} \walk{v_0}{G}{v_n} of \emph{length} $n$ in a graph $G$
is a sequence $v_0,\ldots,v_n$ of nodes with $n\ge1$ and
$(v_i,v_{i+1})\in E$ for $i=0,\ldots,n-1$. A \emph{shortest path}
between nodes $u$ and $v$ is a path \walk{u}{G}{v} of minimal
length. The \emph{transitive closure} of a graph $G=(V,E)$ is given by
$G^*=(V,E^*)$ with $(u,v)\in E^*$ iff there exists a path
\walk{u}{G}{v}. A \emph{strongly connected component (scc)} of $G$ is
a subset $U\subseteq V$, such that \walk{u}{G^*}{v} exists for every
$u,v\in U$. A \emph{(weakly) connected component (wcc)} is defined
accordingly, ignoring the direction of edges $(u,v)\in E$.

Many observations of network properties can be explained just by the
network's degree distribution~\cite{kolaczyk2009statistical}. It is
therefore important to contrast the observed property to the according
result obtained on a random graph as a \emph{null model} which shares
the same degree distribution. If a single network $G$ is considered, a
corresponding null model $\rewire{G}$ can be obtained by randomly
replacing edges $(u_1, v_1), (u_2, v_2)\in E$ with $(u_1, v_2)$ and
$(u_2, v_1)$, ensuring that these edges were not present in $G$
beforehand. This process is typically repeated a multiple of the graph
edge set's cardinality (see~\cite{maslov2002specificity} for
details). For contrasting comparative observations within pairs of
networks $(G_1, G_2)$, a null model $\shuffle{G}_2$ can be obtained by
permuting the vertex positions within $G_2$ as described
in~\cite{butts2005simple}.

\section{Data Sources}\label{sec:data}
\vspace{-0.5\baselineskip}
\paragraph{\wikipedia \& \wiktionary}
For our analysis we used the official \wikipedia data dump which is
freely available for
download\footnote{\url{http://dumps.wikimedia.org/backup-index.html}}
and considered the English (date: \emph{2012-01-05}), French (\emph{2012-01-17}) and
German (\emph{2011-12-12}) version separately. We additionally used the
categorization links of the affiliated \wiktionary project (English,
French and German \emph{2012-06-06}), also available
for download.
\vspace{-0.5\baselineskip}
\paragraph{\twitter}
As an additional source for user generated data we considered the
microblogging service \twitter. Using \twitter, each user publishes
short text messages (called ``\emph{tweets}''). We used the data set
introduced in~\cite{yang2011patterns} which comprises 476,553,560
tweets from 17,069,982 users, collected 2009/06 until
2009/12\vspace{-0.5\baselineskip}
\paragraph{Given Names}
Some effort was made to build up a comprehensive list of given
names. In a semi-automatic way, a list of more than 30,000 names was
collected. During the first months of the \nameling's live time,
additional names were proposed by users of the system, yielding a list
of 36,434 given names.  \vspace{-0.5\baselineskip}
\paragraph{Cities} 
As an example for entities with an obvious ad hoc notion of
relatedness (namely the geographical distance), we considered cities
with a population above 1,000. A corresponding data set which also
comprises corresponding geolocations is freely available for
download\footnote{\url{http://www.geonames.org/}}. We eliminated all
cities with ambiguous names, resulting in a list of 101,667 city
names.

\section{Co-occurrence Networks}\label{sec:networks}
The present work's initial motivation was to find relations among
given names based on user-generated content in the social web. The
most basic relation among such entities can be observed when they
occur together within a given atomic context. In case of \wikipedia,
we counted such \emph{co-occurrences} based on sentences and for
\twitter based on tweets. We thus obtain for each considered entity
type $I\in\set{N,C}$ (given names and city names, respectively) and
data source $S\in\set{\text{EN},\text{DE},\text{FR},\text{Twitter}}$
(English, German and French \wikipedia as well as \twitter) an
undirected weighted graph $G^I_S=(V^I_S,E^I_S)$ where $V^I_S$ denotes
the subset of all observed entities of type $I$ within $S$ and for
entities $u, v$ exists an edge $(u,v)\in E^I_S$ with weight $c$, if
$u$ and $v$ co-occurred in exactly $c$ contexts.

For example, the given names ``\emph{Peter}'' and ``\emph{Paul}''
co-occurred in 30,565 sentences within the English \wikipedia whereas
the city names ``\emph{Kassel}'' and ``\emph{G\"ottingen}''
co-occurred in 630 sentences within the German
\wikipedia. Accordingly, there is an edge $(\text{Peter},
\text{Paul})$ in {\Nen} and an edge $(\text{Kassel},
\text{G\"ottingen})$ in \Cde respectively with corresponding edge
weights.

\subsection{High Level Statistics}\label{sec:statistics}
Table \ref{tab:networks:general} summarizes the high level statistics
for all considered co-occurrence networks. As one would expect, all
networks contain a giant connected
component~\cite{newman2003structure} which almost cover the whole
corresponding node sets. The networks obtained from the English
\wikipedia are the most densely connected network for given names
whereas the French \wikipedia yields the most densely connected
network for city names. Networks obtained from \twitter are least
densely connected.
\begin{table}\vspace{-2\baselineskip}
  \centering
  \caption{High level statistics for all Co-occurrence networks.}
  \begin{tabular}{l|r|r|r|r|r}
             &  \multicolumn{1}{c|}{$|V|$}  
             &  \multicolumn{1}{c|}{$|E|$}         & density   &   \#wcc  & largest wcc \\\hline\hline
    \Nen     &  $27,121$ &  $24,461,988$    &  $0.067$  &   $2$    &   $27,119$  \\\hline
    \Nde     &  $25,032$ &  $13,172,603$    &  $0.042$  &   $2$    &   $25,030$  \\\hline
    \Nfr     &  $25,237$ &  $17,446,139$    &  $0.055$  &   $1$    &   $25,237$  \\\hline
    \NTweets &  $25,902$ &   $6,634,551$    &  $0.020$  &   $1$    &   $25,902$  \\\hline\hline
    \Cen     &  $66,060$ &  $31,015,531$    &  $0.014$  &  $22$    &   $66,014$  \\\hline
    \Cde     &  $58,422$ &  $15,583,628$    &  $0.009$  &  $30$    &   $58,335$  \\\hline
    \Cfr     &  $55,981$ &  $30,786,844$    &  $0.020$  &  $26$    &   $55,927$  \\\hline
    \CTweets &  $42,512$ &   $2,651,510$    &  $0.003$  &  $46$    &   $42,419$  \\\hline
  \end{tabular}
  \label{tab:networks:general}
\end{table}

\todo{Degree Distribution}

\subsection{Inter-Network Analysis}\label{sec:internetwork}
Considering the co-occurrence networks presented above, the question
whether and to which extent these networks are related naturally
arises. 

As a first indicator, we considered basic vertex centrality metrics,
namely \emph{degree centrality} and \emph{eigenvector centrality} as
well as the ``\emph{popularity}'' of an entity, that is, its global
frequency within the corresponding corpus. Please note that we can
directly compare centrality scores for nodes within a family of
networks (given names and city names respectively), as the vertex sets
of these networks are drawn from the same population.

\begin{figure}
  \centering
  \includegraphics[width=0.31\linewidth]{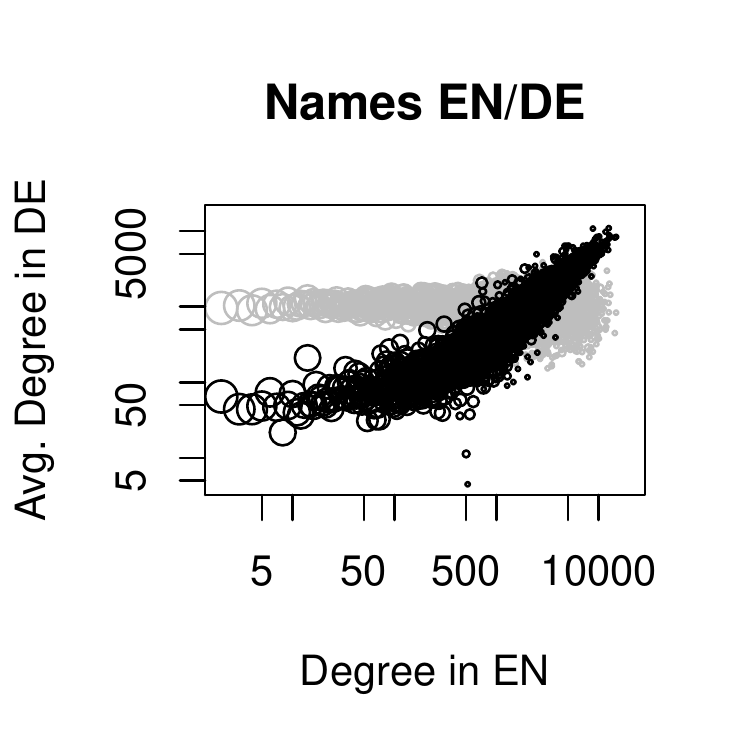}
  \includegraphics[width=0.31\linewidth]{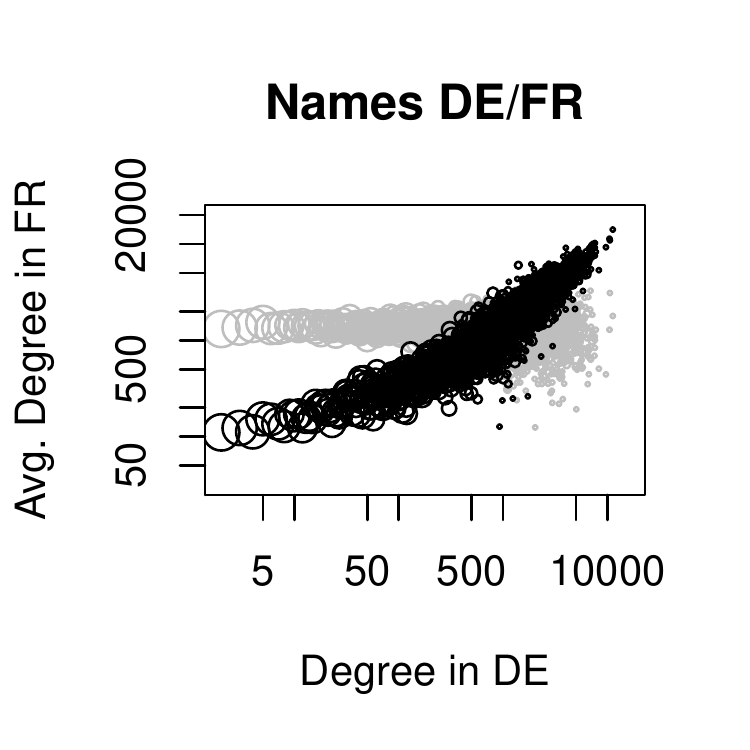}
  \includegraphics[width=0.31\linewidth]{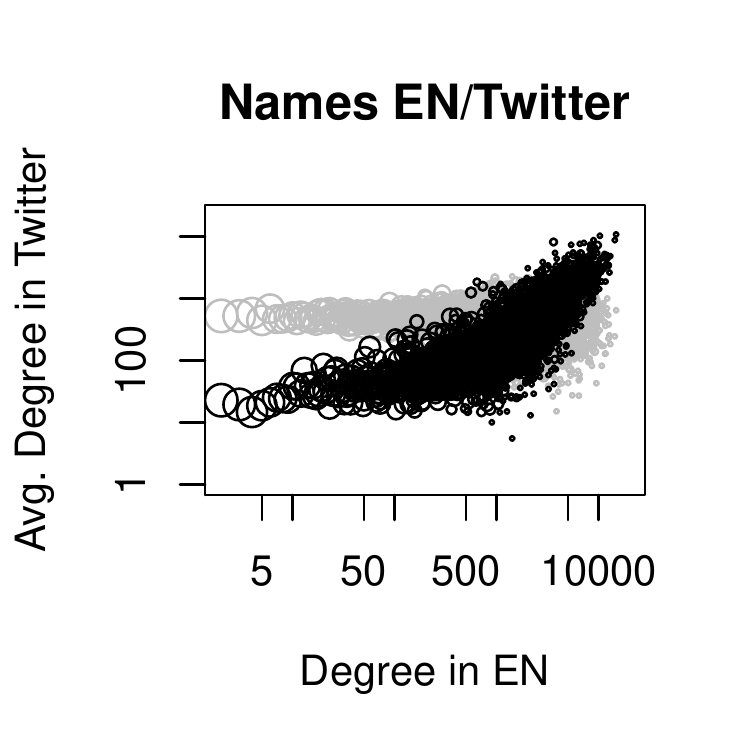}\\
  \includegraphics[width=0.31\linewidth]{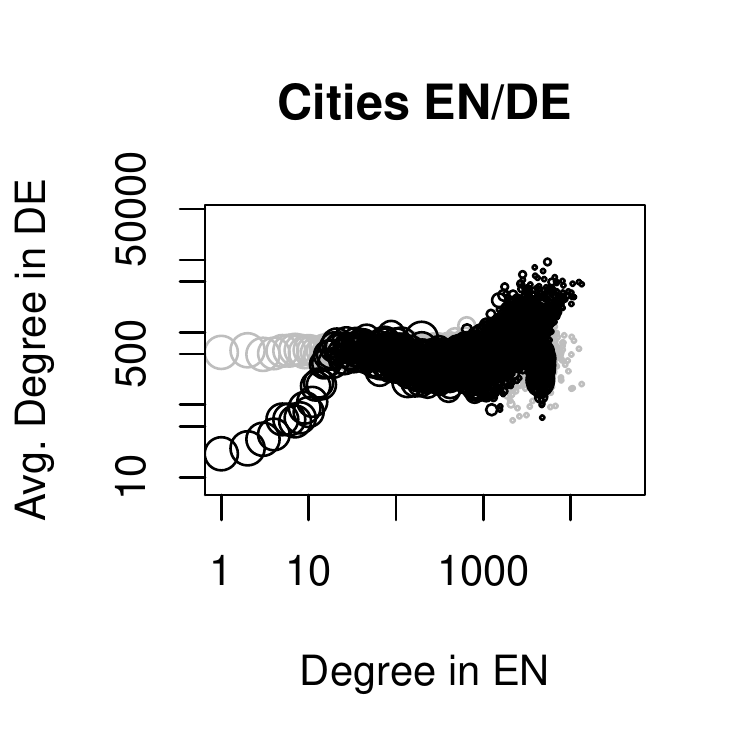}
  \includegraphics[width=0.31\linewidth]{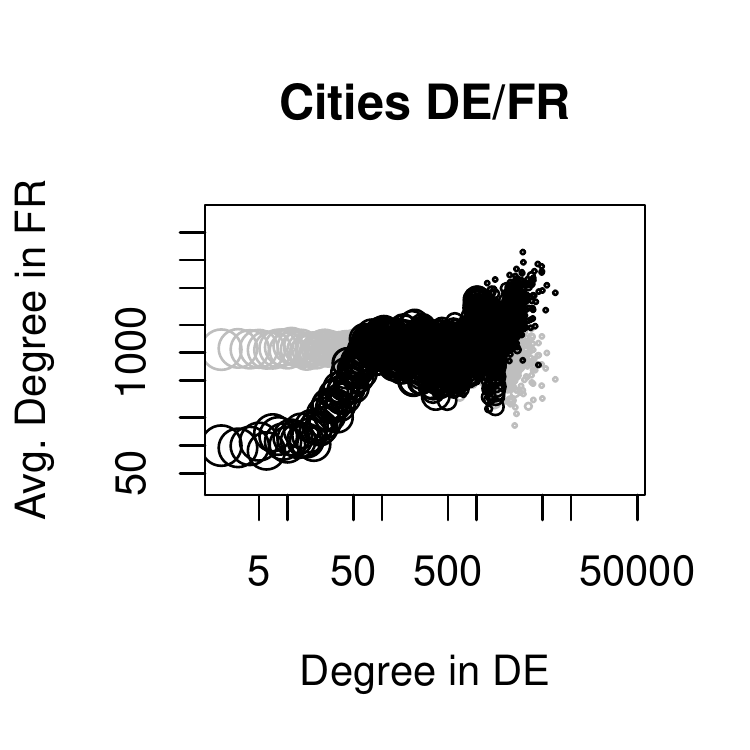}
  \includegraphics[width=0.31\linewidth]{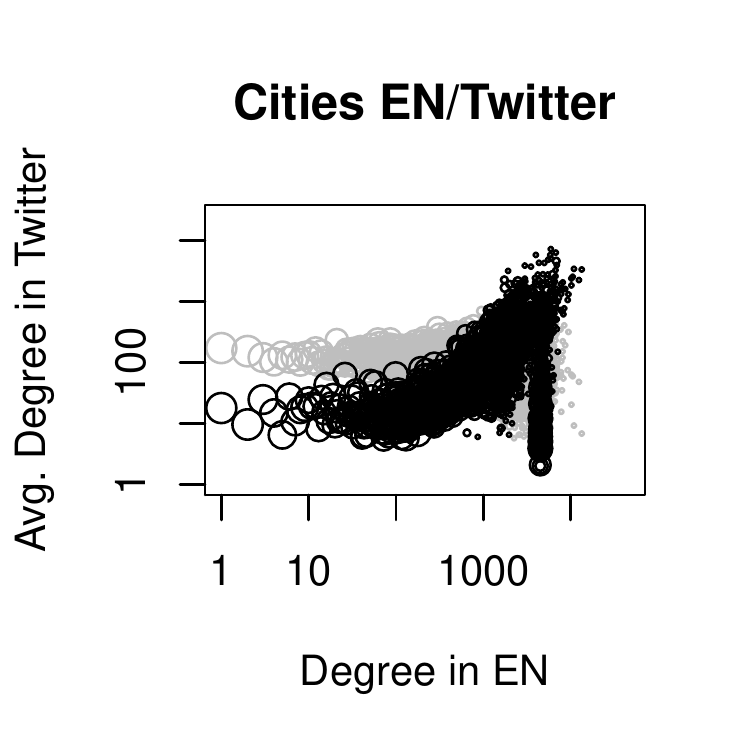}
\comments{
  \includegraphics[width=0.31\linewidth]{results/indc/names-indc-en_de}
  \includegraphics[width=0.31\linewidth]{results/indc/names-indc-de_fr}
  \includegraphics[width=0.31\linewidth]{results/indc/names-indc-en_twitter}\\
  \includegraphics[width=0.31\linewidth]{results/indc/cities-indc-en_de}
  \includegraphics[width=0.31\linewidth]{results/indc/cities-indc-de_fr}
  \includegraphics[width=0.31\linewidth]{results/indc/cities-indc-en_twitter}
}
\caption{Degree centrality in co-occurrence networks derived from the
  English (EN), French (FR) and German (DE) \wikipedia, where results
  obtained from corresponding null models are depicted in gray.}
  \label{fig:internetwork:degree}
\end{figure}

Figure~\ref{fig:internetwork:degree} exemplarily shows a pairwise
comparison of the degree centrality within different networks
$G_1,G_2$. To reduce noise, we calculated for all names having a
degree of $k$ in $G_1$ the average node degree in $G_2$ and scaled the
point size logarithmically with the number of corresponding
observations. The top left plot, \eg shows that in average a given
name with a degree of 50 in the English \wikipedia has a degree of
comparable magnitude in the German \wikipedia. Due to the underlying
heavy tailed distributions we plotted in a logarithmic
scale. \todo{Why not scale point size with variance?} To rule out
effects induced by the graphs' degree distributions, we considered for
each pair $G,G'$ of networks a corresponding \emph{null model}
$\shuffle{G}'$ (see Sec.~\ref{sec:preliminaries}) where effectively
the degree distribution of $G'$ is fixed but the vertices are permuted
randomly. The results for the null models are averaged for repeated
calculations and depicted in gray.

As a general trend, positive correlations for the degree centrality
can be observed in all networks for given names, though less
pronounced for the \twitter based network and for lower vertex
degrees but significantly deviating from correlations obtained from a
corresponding null model.

For the city name networks, positively correlated trends can only be
observed for lower degree nodes in the \wikipedia based networks. For
the \twitter based network the result is comparable with the given
names networks. Please note the significant cluster of nodes with high
degree centrality in the English \wikipedia and low centrality scores
for the other networks. Manual inspection showed that these are indeed
results of corresponding distinct city names and not names with common
words. These outliers can not be explained just by analyzing the
network structure and therefor the word contexts within the corpora
must be considered which is out of the present work's scope.

\begin{figure}
  \centering
  \includegraphics[width=0.45\linewidth]{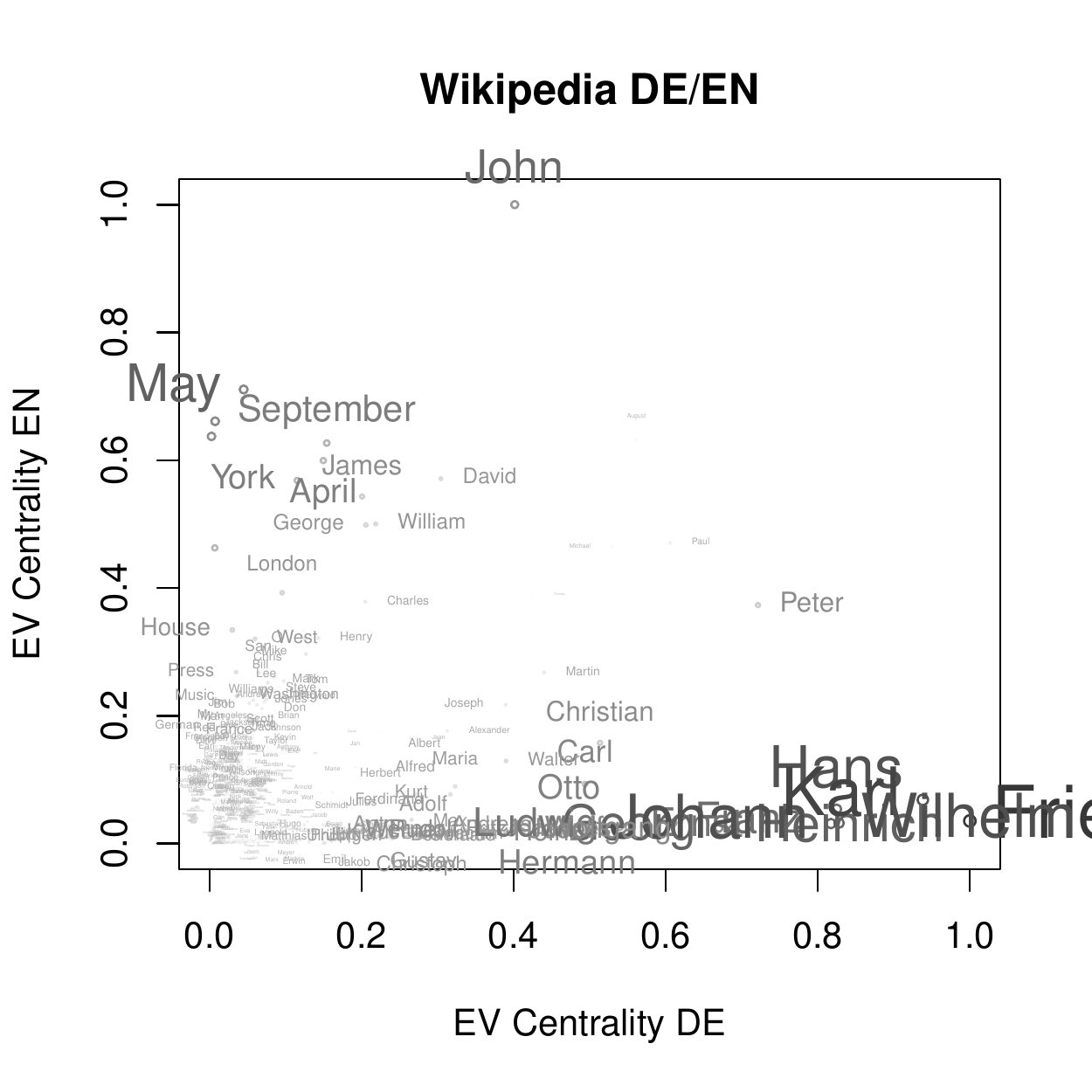}
  \includegraphics[width=0.45\linewidth]{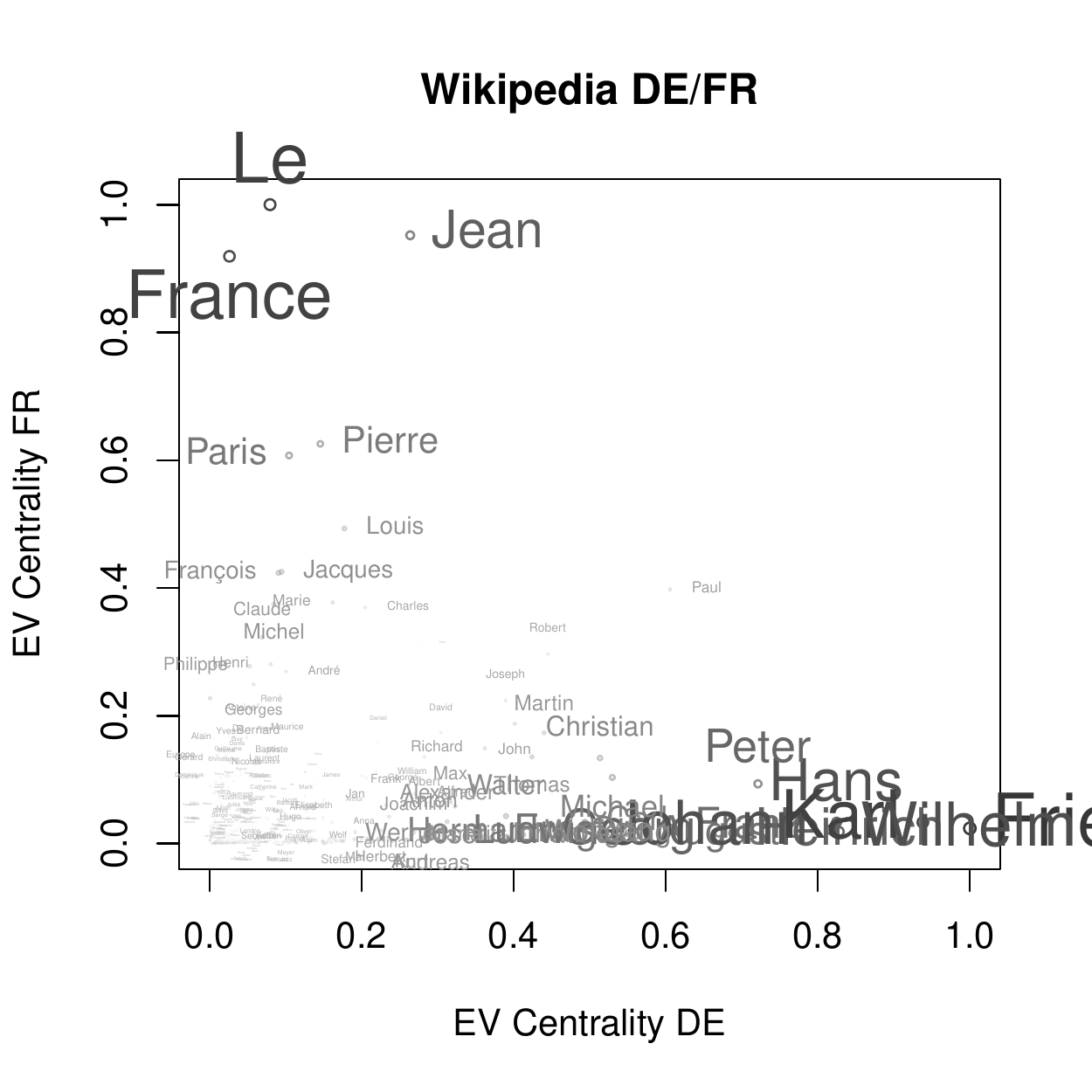}
  \caption{Pairwise comparison of eigenvector centrality for
    co-occurrence networks of given names based on \wikipedia in
    English, French and German.}
  \label{fig:internetwork:ev:names}
\end{figure}

In contrast to the degree centrality, the eigenvector centrality
appears to reveal distinct trends for given names within the
corresponding co-occurrence networks. Figure
\ref{fig:internetwork:ev:names} exemplarily shows the comparative
plots for eigenvector centrality within pairs of given name
networks. In both cases, the lower right area is (by trend) populated
with classic German names whereas the upper left area is populated by
English and French names, respectively. These language specific
characteristics of the eigenvector centrality can be exploited for
automatically classifying given names according to their cultural
background.

\begin{figure}
  \begin{minipage}[t]{0.6\linewidth}
    \vspace{0pt}\includegraphics[scale=0.5]{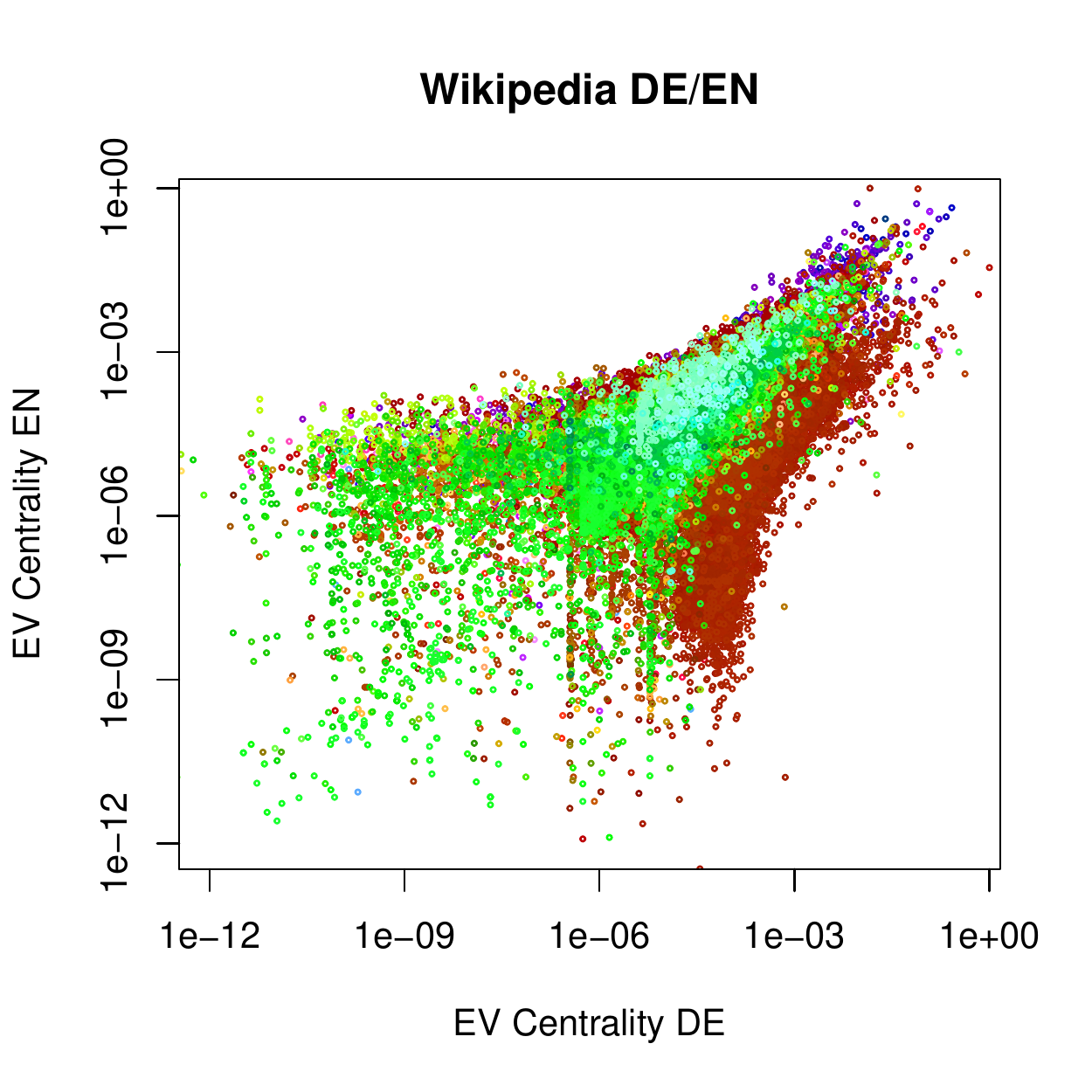}
  \end{minipage}
  \begin{minipage}[t]{0.35\linewidth}
    \vspace{27pt}\includegraphics[scale=0.5]{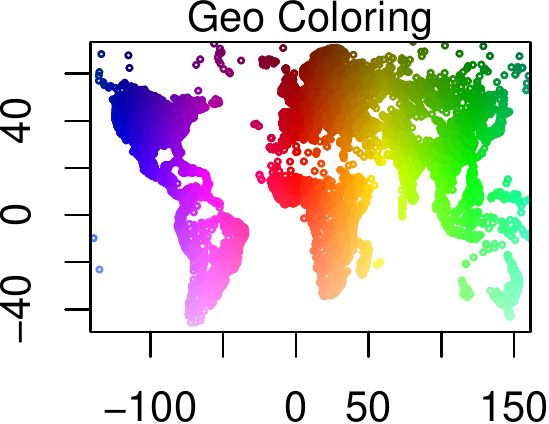}
    \includegraphics[scale=0.5]{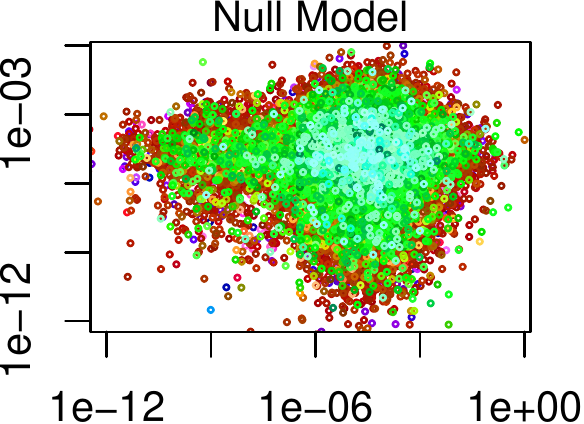}
  \end{minipage}
  \caption{Pairwise comparison of eigenvector centrality for
    co-occurrence networks of city names based on \wikipedia in
    English and German.}
  \label{fig:internetwork:ev:cities}
\end{figure}

For the city names networks, the eigenvector centrality exhibits only
very sparse distinct language specific trends which are dominated by
city names which coincide with common words of the respective
language, as for example ``\emph{England}'', ``\emph{Collage}'' and
``\emph{Church}'' for English and ``\emph{Das}'', ``\emph{Die}'',
``\emph{Band}'' for German. Most of the centrality scores are
clustered together and show a significantly correlated trend in the
corresponding log-scale plot in
Fig.~\ref{fig:internetwork:ev:cities}. For visualizing the
geographical reference of the denoted cities, we colored each point
according to the respective geographic location, where latitude and
longitude are used to select a color within the HSL color space (see
the top right earth globe projection in
Fig.~\ref{fig:internetwork:ev:cities}). Please note that points are
plotted ordered according to the corresponding longitude value for
unifying the effect of covered areas. Comparing with the null model
(obtained by comparing $G^C_{\text{DE}}$ with
$\shuffle{G}^C_{\text{EN}}$), Fig.~\ref{fig:internetwork:ev:cities}
reveals a correlated trend for the eigenvector centrality of city
names in the different language specific editions of \wikipedia and
points towards an interrelation of the geographic location of a city
and its position within the co-occurrence networks. We will
investigate this interrelation more detailed in
Sec.~\ref{sec:similarities}.
\subsection{Inter-Network Correlation Test}\label{sec:internetwork:qap}
For a more formalized analysis, we assess the network interrelation in
terms of the correlation of the corresponding adjacency matrices by
applying the quadradic assignment procedure (\qap)
test~\cite{butts2008social,butts2005simple}.

For given graphs $G_1=(V_1, E_1)$ and $G_2=(V_2,E_2)$ with $U\coloneqq
V_1\cap V_2\ne\emptyset$ and adjacency matrices $A_i$ corresponding to
${G_i}_{|U}$ ($G_i$ reduced to the common vertex set $U$, see
Sec. \ref{sec:preliminaries}), the graph \emph{covariance} is given by
\[
\cov(G_1,G_2) \coloneqq
\frac{1}{n^2-1}\sum_{i=1}^{n}\sum_{j=1}^{n}(A_1[i,j]-\mu_1)(A_2[i,j]-\mu_2)
\]
where $n\coloneqq\size{U}$ and $\mu_i$ denotes $A_i$'s mean
($i=1,2$). Then $\var(G_i)\coloneqq\cov(G_i, G_i)$ leading to the
graph correlation
\[
\rho(G_1,G_2)\coloneqq
\frac{\cov(G_1,G_2)}{\sqrt{\var(G_1)\var(G_2)}}.
\]
The \qap test compares the observed graph correlation $\rho_0$ to the
distribution of resulting correlation scores obtained on repeated
random row/column permutations of $A_2$. The fraction of permutations
$\pi$ with correlation $\rho^\pi\ge\rho_o$ is used for assessing the
significance of an observed correlation score $\rho_o$. Intuitively,
the test determines (asymptotically) the fraction of all graphs with
the same structure as $G_{2|U}$ having at least the same level of
correlation with $G_{1|U}$.

Table \ref{tab:qap} shows the pairwise correlation scores for all
considered networks. Consistent with our preceding observations, the
\wikipedia-based co-occurrence shows the strongest correlations,
significantly more pronounced for given names. For assessing the
significance of the observed correlations, we repeatedly calculated
the pairwise correlations on 1,000 corresponding randomly generated
null models. For any pair of the considered networks, every randomly
generated null model showed much lower correlation scores ($<10^{-3}$),
which indicates according to~\cite{butts2008social} statistical
significance.

We conclude that the co-occurrence networks structurally correlate,
more pronounced though for given names than for city
names. Nevertheless, language specific deviations exist. For
discovering relations on named entities the corresponding language
should therefore be considered. In the next section we will
investigate, how relations can be extracted from the co-occurrence
networks and how these relations correlate with natural notions of
relatedness.
\comments{Figure~\ref{fig:internetwork:qap} shows exemplary distribution
of the resulting correlation scores which are consistently lower by
magnitude then the observed correlations in the original graphs, which
suggests that even the low correlation scores for the \twitter based
networks are significant.}
\begin{table}[ht]\centering
  \caption{Pairwise graph correlation observed in the co-occurrence graphs.}
  \begin{minipage}{0.45\linewidth}
    \begin{tabular}{l|r|r|r|r}
      &  \Nen     &  \Nde      &  \Nfr     &  \NTweets \\\hline\hline
      \Nen     &  \multicolumn{1}{c|}{-}        &  $0.406$   &  $0.332$  &  $0.049$  \\\hline
      \Nde     &  \multicolumn{1}{c|}{-}        &  \multicolumn{1}{c|}{-}         &  $0.303$  &  $0.020$  \\\hline
      \Nfr     &  \multicolumn{1}{c|}{-}        &  \multicolumn{1}{c|}{-}         &  \multicolumn{1}{c|}{-}        &  $0.015$  \\\hline
      \NTweets &  \multicolumn{1}{c|}{-}        &  \multicolumn{1}{c|}{-}         &  \multicolumn{1}{c|}{-}        &  \multicolumn{1}{c}{-}        \\
    \end{tabular}
  \end{minipage}
  \begin{minipage}{0.45\linewidth}
    \begin{tabular}{l|r|r|r|r}
      &  \Cen    &  \Cde       &  \Cfr      &  \CTweets \\\hline\hline
      \Cen     &  \multicolumn{1}{c|}{-}       &  $0.119$    &  $0.180$   &  $0.067$  \\\hline
      \Cde     &  \multicolumn{1}{c|}{-}       &  \multicolumn{1}{c|}{-}          &  $0.135$   &  $0.025$  \\\hline
      \Cfr     &  \multicolumn{1}{c|}{-}       &  \multicolumn{1}{c|}{-}          &  \multicolumn{1}{c|}{-}         &  $0.040$  \\\hline
      \CTweets &  \multicolumn{1}{c|}{-}       &  \multicolumn{1}{c|}{-}          &  \multicolumn{1}{c|}{-}         &  \multicolumn{1}{c}{-}        \\
    \end{tabular}
  \end{minipage}
  \label{tab:qap}
\end{table}

\comments{
All networks within \bibs show a consistent level
of correlation with a significant peak for the explicit \Friend- and
\Group networks. Considering the results for the networks obtained
from \flickr it is worth noting that, though low in magnitute, the
Favorite-Graph shows a significant higher correlation with the
Contact-Graph than the other pairs of networks do.
}

\comments{
\begin{figure}
  \centering
  \includegraphics[width=0.3\linewidth]{results/internetwork/qap/de-en-1000}
  \includegraphics[width=0.3\linewidth]{results/internetwork/qap/de-fr-1000}
  \includegraphics[width=0.3\linewidth]{results/internetwork/qap/en-fr-1000}\comments{\\
  \includegraphics[width=0.3\linewidth]{results/internetwork/qap/en-twitter-1000}
  \includegraphics[width=0.3\linewidth]{results/internetwork/qap/de-twitter-1000}
  \includegraphics[width=0.3\linewidth]{results/internetwork/qap/fr-twitter-1000}}
  
  \caption{Test statistics for the QAP correlation test.}
  \label{fig:internetwork:qap}
\end{figure}
}
\section{Mining for Relations from the Social Web}\label{sec:task}\enlargethispage*{4\baselineskip}
In the previous section, two kinds of co-occurrence networks were
introduced, one for given names and the other for city names. These
networks were structurally analyzed and compared, giving insights into
immanent properties and correlations among different networks. Keeping
in mind the initial motivation for the present work, namely the
recommendation of given names based on data from the social web, we
now focus on the question, whether the co-occurrence networks from
Section~\ref{sec:networks} give raise to a notion of relatedness which
implies relationships the user might be interested in.

For evaluating different similarity metrics based on the co-occurrence
networks, we need a ``reference'' notion of relatedness for the
considered entities to be used as ``ground truth''. For cities, the
geographic distance is a natural candidate. For given names, such
generally accepted reference relation is less obvious. We therefor
apply the approach of using an external data source which we assume as
a valid ``gold standard''. We argue that the categories assigned to
names in \wiktionary are a good basis, as they are manually assigned
and have a direct connection to concepts users associate with given
names (such as gender and cultural context). We finally chose cosine
similarity (see Sec.~\ref{sec:metrics}) for calculating a reference
similarity score, which is broadly accepted for various
applications. \comments{We also calculated the Jaccard coefficient for assessing
similarity, but due to the strong correlation between cosine
similarity and Jaccard coefficient, no significant difference could be
observed. }For simplicity we restrict our analysis in this chapter to
the English \wikipedia.

In the following, we will first introduce different similarity
functions for calculating similarity of named entities based on
corresponding co-occurrence networks. We will than compare these
similarity functions for given names and city names, respectively,
with the corresponding gold standard relations described above.
\todo{future work: Of course, whether or not the considered similarity
  scores coincide with a concept of relatedness a user has in mind can
  ultimately only be answered by evaluating against usage statistics
  and user feedback in a running system, such as, \eg the \nameling.
}

\subsection{Vertex Similarities}\label{sec:similarities}\label{sec:metrics}
Similarity scores for pairs of vertices based only on the surrounding
network structure have a broad range of applications, especially for
the link prediction task~\cite{libennowell2007linkprediction}. In the
following we present all considered similarity functions, following
the presentation given in \cite{desa2011supervised} which builds on
the extensions of standard similarity functions for weighted
networks from~\cite{murata2007prediction}.
\comments{
\paragraph{Common Neighbors (\CN)}
The \CN metric simply counts common neighbors for pairs of vertices:
\[
\CN(x,y)\coloneqq\size{\Gamma(x)\cap\Gamma(y)}
\]
Although simple, the \CN is widely adopted in most online social
networks and shows good performance for predicting
links~\cite{libennowell2007linkprediction} and is straightforward
extended to weighted networks:
\[
\wCN(x,y)\coloneqq\sum_{z\in\Gamma(x)\cap\Gamma(y)}w(x,z)+w(y,z)
\]
}
The \emph{Jaccard coefficient} measures the fraction of common neighbors:
\[
\JC(x,y)\coloneqq \frac{\size{\Gamma(x)\cap\Gamma(y)}}{\size{\Gamma(x)\cup\Gamma(y)}}
\]
The Jaccard coefficient is broadly applicable and commonly used for
various data mining tasks. For weighted networks the Jaccard
coefficient becomes:
\[
\wJC(x,y)\coloneqq \sum_{z\in\Gamma(x)\cap\Gamma(y)}
  \frac{w(x,z)+w(y,z)}
       {\sum_{a\in\Gamma(x)}w(a,x)+\sum_{b\in\Gamma(y)}w(b,y)}
\]
\comments{
\paragraph{Preferential Attachment (\PA)}
In the context of social networks, preferential attachment describes
the effect in growing networks, that high degree vertices tend to
connect with other high degree
vertices~\cite{newman2003structure}. These effects can be exploited
for predicting the probability of future links~\cite{newman2001clustering}:
\[
\PA(x,y)\coloneqq\size{\Gamma(x)}\cdot\size{\Gamma(y)}
\]
For weighted networks the \PA can be extended as:
\[
\wPA(x,y)\coloneqq\sum_{a\in\Gamma(x)}w(a,x)\cdot\sum_{b\in\Gamma(y)}w(b,y)
\]
}
The \emph{resource allocation index} \RA~\cite{zhou2009predicting} captures the
intuition that for two nodes $x$ and $y$, the importance of a common neighbor
$z$ to their relatedness depends on how ``exclusive'' $z$ connects $x$ with $y$:
\[
\RA(x,y)\coloneqq\sum_{z\in\Gamma(x)\cap\Gamma(y)}\frac{1}{\size{\Gamma(z)}}
\]
The \RA for weighted networks is given by
\[
\wRA(x,y)\coloneqq\sum_{z\in\Gamma(x)\cap\Gamma(y)}
   \frac{w(x,z)+w(y,z)}
        {\sum_{c\in\Gamma(z)}w(z,c)}.
\]%
Similar to \RA, the \emph{Adamic-Adar} coefficient captures the exclusiveness
of common neighbors, though respecting underlying power distributions:
\[
\AC(x,y)\coloneqq\sum_{z\in\Gamma(x)\cap\Gamma(y)}\frac{1}{\log(\size{\Gamma(z)})}
\]
For weighted networks, the Adamic-Adar coefficient is defined as
\[
\wAC(x,y)\coloneqq\sum_{z\in\Gamma(x)\cap\Gamma(y)}
   \frac{w(x,z)+w(y,z)}
        {\log(1+\sum_{c\in\Gamma(z)}w(z,c))}.
\]
%
The \emph{cosine similarity} measures the cosine of the angle between
the corresponding rows of the adjacency matrix, which for a unweighted
graph can be expressed as
\[
\CS(x,y) \coloneqq \frac{\size{\Gamma(x)\cap\Gamma(y)}}{\sqrt{\size{\Gamma(x)}}\cdot\sqrt{\size{\Gamma(y)}}},
\]
and for a weighted graph is given by
\[
\wCS(x,y) \coloneqq 
\sum_{z\in\Gamma(x)\cap\Gamma(y)}
\frac
{
  w(x,z)w(y,z)
}
{
  \sqrt{
  \sum_{a\in\Gamma(x)}w(x,a)^2}\cdot\sqrt{\sum_{b\in\Gamma(y)}w(y,b)^2}
}.
\]%
%
The vertex similarity introduced in~\cite{leicht2005vertex} measures
the observed number of common neighbors relative to the overlap expected in
a corresponding random graph:
\[
\NM(x,y) \coloneqq \frac{
  \size{\Gamma(x)\cap\Gamma(y)}
}{
  \size{\Gamma(x)}\size{\Gamma(y)}
}
\]

\subsection{Given Names}\label{sec:experiments:names}\enlargethispage*{5\baselineskip}
For obtaining a reference relation on the set of given names, we
collected all corresponding category assignments from \wiktionary. We
thus obtained for each of 10,938 given names a respective binary
vector, where each component indicates whether the corresponding
category was assigned to it (in total 7,923 different categories and
80,726 non-zero entries). As these assignment vectors are very sparse,
we counted for each name the number of name pairs with a non-zero
similarity score, to ensure that a relevant similarity metric is
induced. Indeed, more than 90\% of the names had more than one
hundred ``similar'' names.

\begin{figure}
  \centering
  \includegraphics[width=0.45\linewidth]{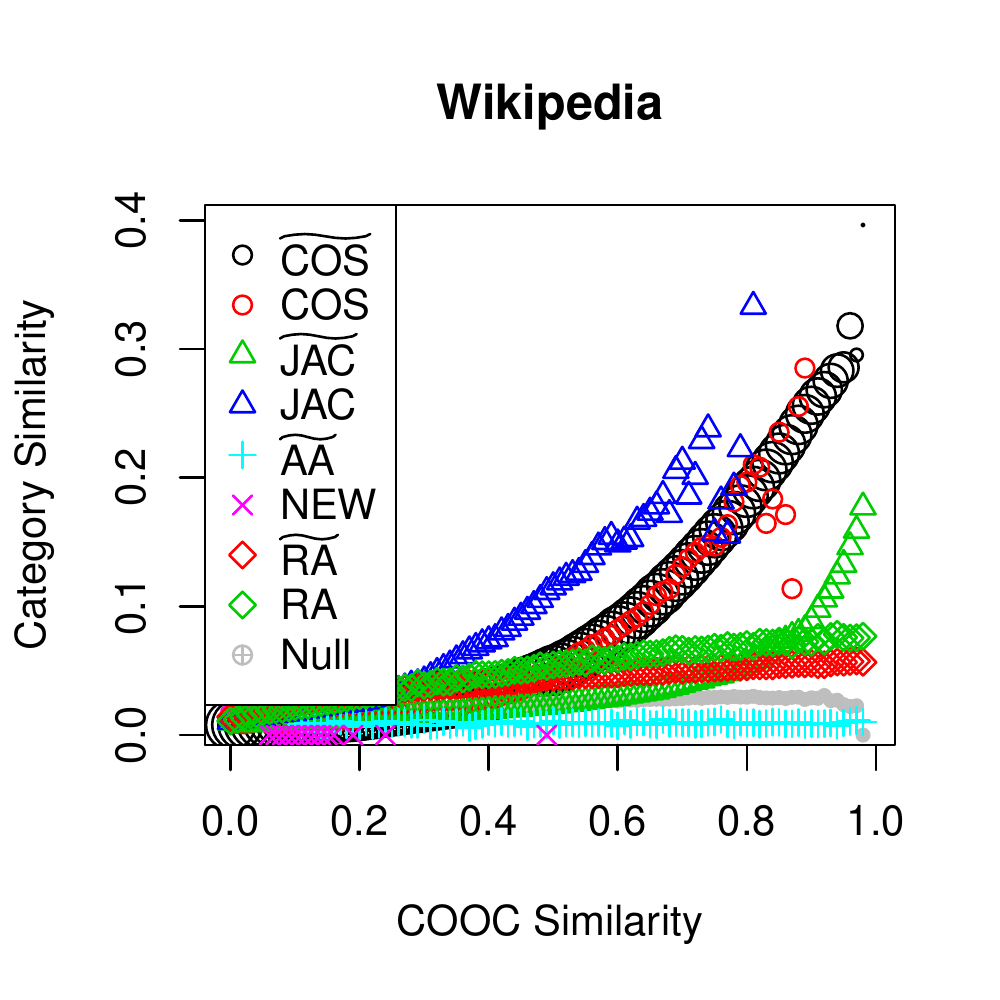}
  \includegraphics[width=0.45\linewidth]{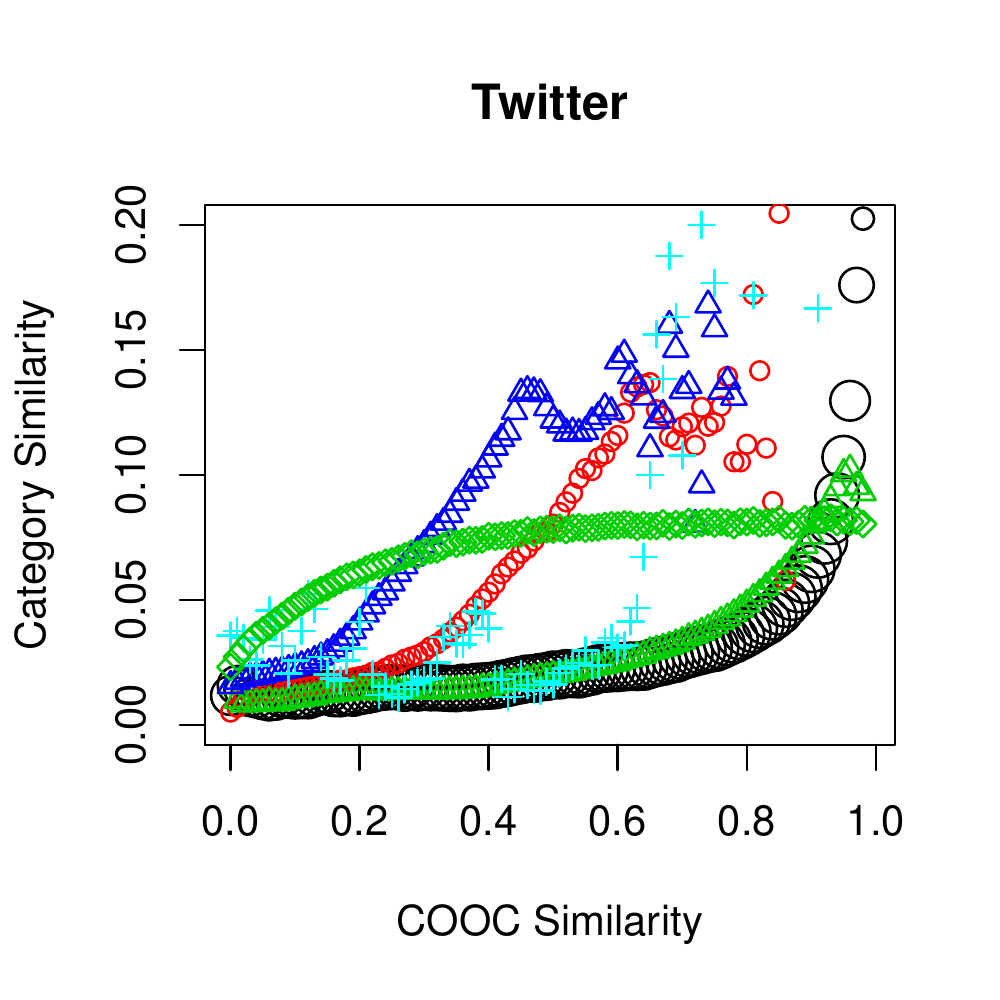}

  \caption{Similarity based on name categories in \wiktionary
    vs. vertex similarity in the \cooc-networks (weighted and
    unweighted).}
  \label{fig:similarities:semsim}
\end{figure}

For any pair $u,v$ of names in the co-occurrence network which have a
category assignment, we calculated the cosine similarity $\CS(u,v)$
based on the respective category assignment vectors as well as any of
the similarity metrics $s(u,v)$ described in section
\ref{sec:similarities}. As the number of data points $(\CS(uv),
S(u,v))$ grows quadratically with the number of names, we grouped the
co-occurrence based similarity scores in 1,000 equidistant bins and
calculated for each bin the average cosine similarity based on
category assignments. Figure~\ref{fig:similarities:semsim} shows the
results for \wikipedia and \twitter separately.

Notably all but Adamic-Adar and \NM capture a positive correlation
between similarity in the co-occurrence network and similarity between
category assignments to names. But significant differences between the
underlying co-occurrence networks and the applied similarity functions
can be observed. As for \wikipedia, the weighted cosine similarity
performs very well, firstly in showing a steep slope and secondly in
exhibiting a stable monotonous curve progression. The unweighted
Jaccard coefficient also shows an even more pronounced linear
progression, but is less stable for higher similarity scores whereas
the weighted Jaccard coefficient shows a higher correlation with the
reference similarity for high similarity scores.

As for \twitter, no globally best matching similarity score can be
found. Each of the similarity functions shows good progression only
in parts. Considering only cosine similarity and the Jaccard
coefficient, we see that both in the unweighted case show higher
correlations with the semantic similarity for mid range similarity
scores, whereas in the weighted case, both exhibit higher correlations
for higher similarity scores. 

\subsection{City Names}\label{sec:experiments:cities}\enlargethispage*{4\baselineskip}
We conducted the same experiment as in section
\ref{sec:experiments:names} for city names, using the geographical
distance of corresponding pairs of cities as a reference relation. As
we only considered cities with a unique name in the data set (see
Sec.~\ref{sec:data}) and each city has a distinct geographical
location associated, we thus obtained a dense reference relation with
explicit real world semantics associated.

\begin{figure}
  \centering
  \includegraphics[width=0.49\linewidth]{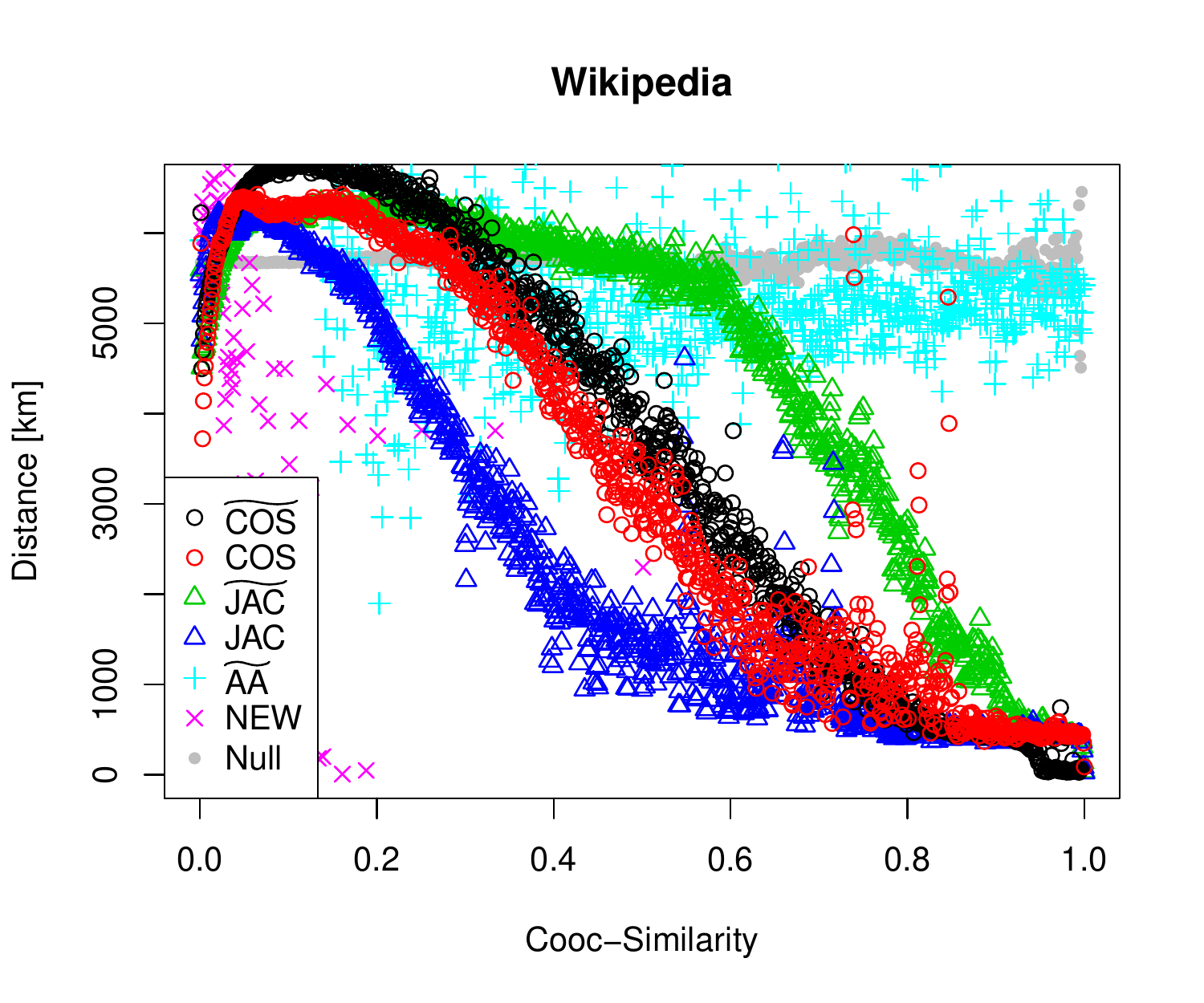}
  \includegraphics[width=0.49\linewidth]{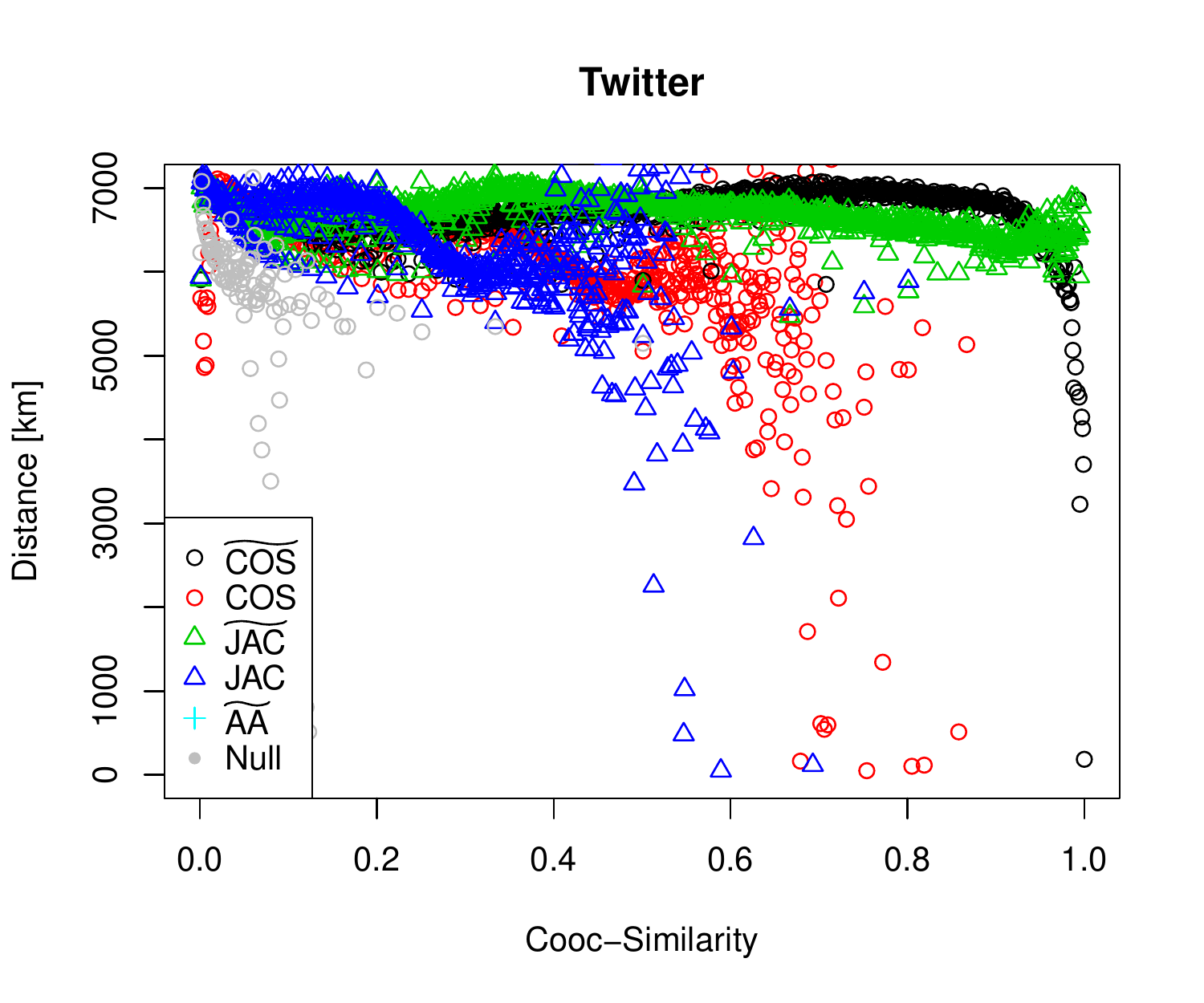}

  \caption{Geographic distance between cities versus vertex
    similarites in the \cooc-networks (weighted and unweighted).}
  \label{fig:similarities:geodist}
\end{figure}

We calculated for each pair $u,v$ of city names the geographical
distance $d(u,v)$ and similarity $s(u,v)$ in the co-occurrence
networks (see Sec. \ref{sec:similarities}). As the number of data
points grows quadratically with the number of city names, we grouped
the co-occurrence based similarity scores in 1,000 equidistant bins
and calculated for each bin the average geographical
distance. Figure~\ref{fig:similarities:geodist} shows the resulting
plots for all considered similarity functions on the \wikipedia and
\twitter-based co-occurrence networks separately.

\enlargethispage*{2\baselineskip}
Considering the results obtained on \wikipedia, cosine similarity and
the Jaccard coefficient show a strikingly high correlation with the
geographical distance. For the cosine similarity and the weighted
Jaccard coefficient, a negative correlation can be observed for
similarity scores $\le 0.2$ which is also present for the unweighted
Jaccard coefficient for low similarity scores $\le 0.05$. We also
counted the number of observations per bin to rule out effects induced
by averaging the geographical distance, but no significant
accumulation of low similarity scores $\le 0.2$ could be observed. We
conclude that low similarity scores in the co-occurrence based
networks are less significant. Both cosine similarity and Jaccard
coefficient show more stable results in the weighted variant, where
cosine similarity shows most significant correlations for mid-range
similarity scores whereas the Jaccard coefficient performs best for
higher similarity scores. For all other similarity metrics, no
correlation can be observed, where the resource allocation index is
excluded for a clearer presentation.

As for \twitter, no significant correlation between structural
similarity in the co-occurrence network and geographical distance can
be observed, despite a very small range around very high similarity
scores of the weighted cosine similarity. The next section
investigates this deviating characteristics in more details.

\subsection{Neighborhood \& Similarity}\label{sec:experiments:simdist}\enlargethispage*{2\baselineskip}
In the preceding sections, the correlation of external reference
measures of semantic relatedness with different similarity functions
in the co-occurrence networks was analyzed. For given names,
correlations could be observed in networks obtained from \wikipedia
and \twitter, whereas for city names, the \wikipedia based analysis
showed astonishing high correlations in contrast to the \twitter based
network, where no significant correlations could be observed.

For further analysis, we considered a very basic measure of
relatedness between two nodes in a network, namely their respective
shortest path distance. We asked, whether names which are direct
neighbors in the co-occurrence graph tend to be more similar than
distant names and whether cities which occur together tend to be
located geographically nearby. That is, for every shortest path
distance $d$ and every pair of nodes $u,v$ with a shortest path
distance $d$, we calculated the average corresponding similarity score
($\CS(u,v)$ and $\JC(u,v)$ for given names and geographic distance
between $u$ and $v$ for city names). To rule out statistical effects,
we repeated for each network $G$ the same calculations on
corresponding null model graphs $\shuffle{G}$.

\begin{figure}
  \centering
  \subfloat[][Given Names]{
    \includegraphics[width=0.24\linewidth]{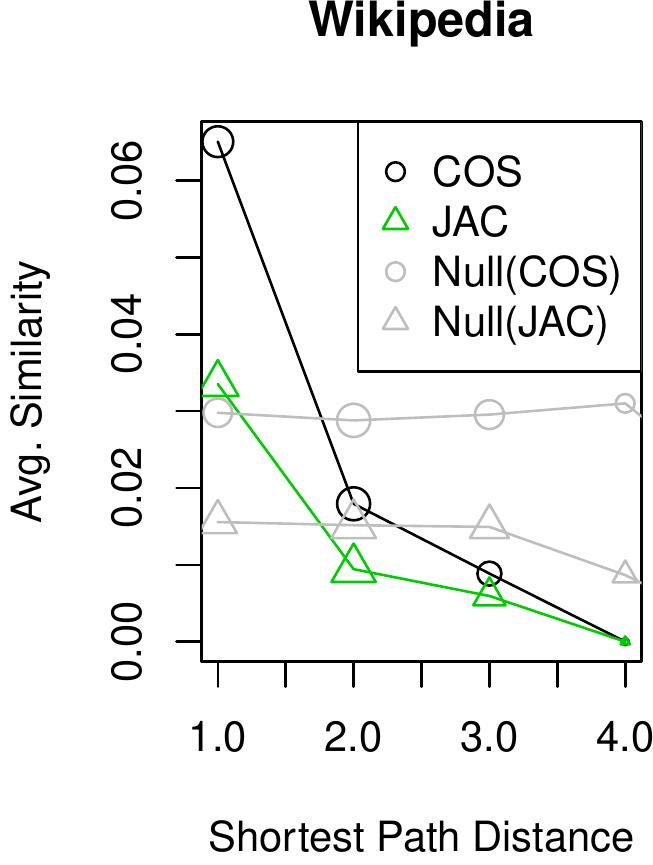}
    \includegraphics[width=0.24\linewidth]{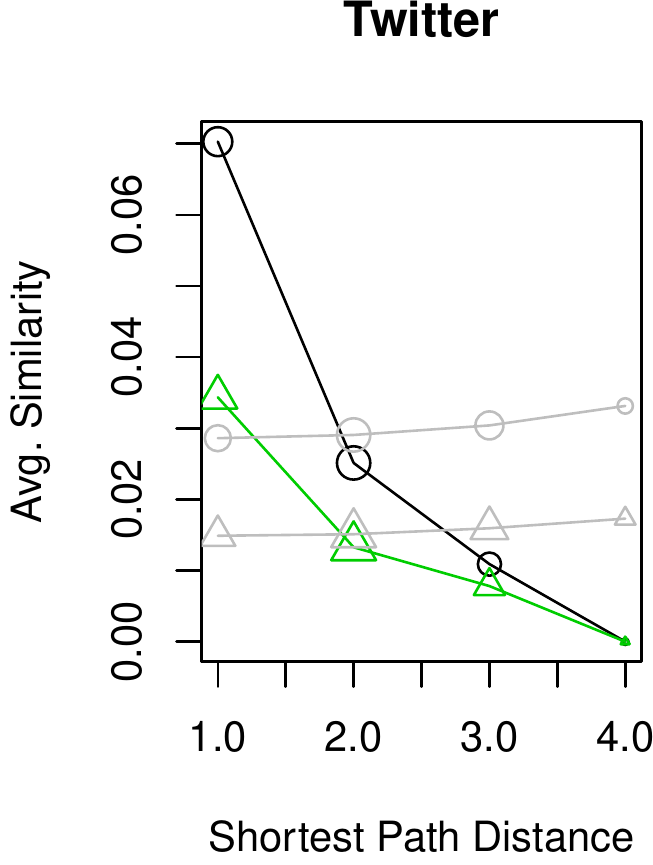}
    \ 
  }
  \subfloat[][City Names]{
    \ 
    \includegraphics[width=0.24\linewidth]{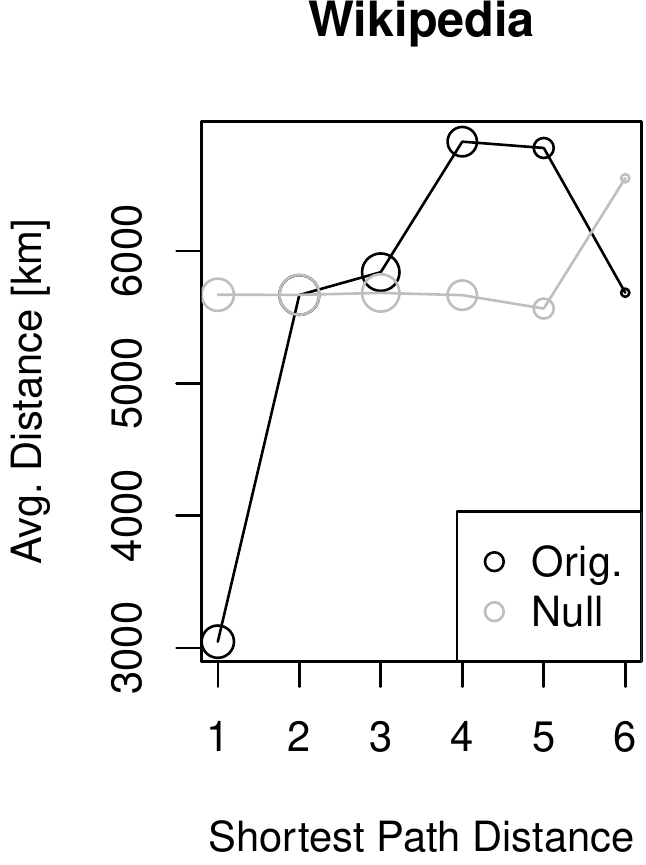}
    \includegraphics[width=0.24\linewidth]{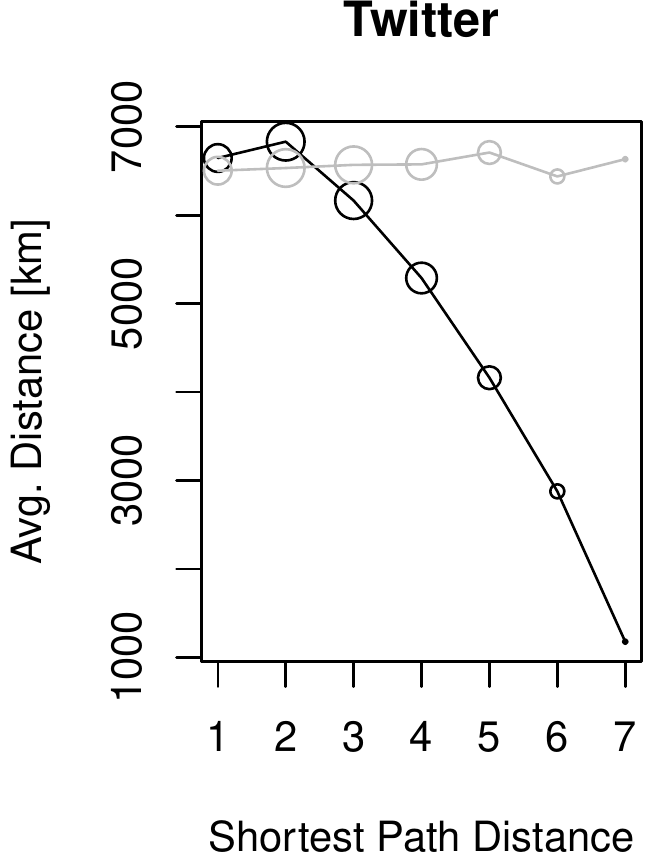}
  }
  \caption{Semantic relatedness vs. shortest path distance in the co-occurrence networks.}
  \label{fig:similarities:simdist}
\end{figure}

Figure~\ref{fig:similarities:simdist} shows the results for given
names and the cosine similarity together with the Jaccard coefficient
as well as for city names and geographical distance. For given names,
in both networks the similarity of node pairs tends to decrease
monotonically with the respective shortest path distance, where direct
neighbors are in average more similar than randomly chosen pairs
(refer to the null model baseline) and pairs at distance two are
already less similar than expected by chance. As for city names, the
\wikipedia based network shows an positive correlation of shortest
path distance with geographical distance, where the deviating behavior
of nodes at distance six is not statistically significant, as only 83
pairs of nodes with corresponding distance exist (in contrast to over
31 million direct neighbors). Most notably, the relationship of
geographical distance and shortest path distance in the \twitter based
network is inverse. Further experimentation for explaining this
deviating semantics are out of the scope of the present work. But it
shows that the semantics induced by co-occurrence in \twitter differs
from the semantics induced by \wikipedia. It explains the difference
in the observed correlations for similarity and geographical distance
of city names in \wikipedia and \twitter based co-occurrence networks
in Sec.~\ref{sec:experiments:cities}, as the considered similarity
functions only depend on the direct neighborhood.

\comments{
\section{Discussion}\label{sec:results}
}
\todo{
  * twitter categories: combination + machine learning?
  * twitter: different semantics?
}
\section{Conclusion \& Future Work}\label{sec:closing}\enlargethispage*{2\baselineskip}
\todo{similarity optimization as a machine learning task} 
With the present work, we introduce the task of discovering
relatedness of given names based on data from the social web. Our
experiments, on the one side, show promising results already for well
known basic approaches, namely co-occurrence based similarity
calculations. On the other side, the presented analysis builds an
experimental framework for analyzing semantics captured by different
co-occurrence networks. The present work already yields results of
practical relevance, underpinned by the success of the \nameling,
which allows users to browse through given names, using the discovered
relations among given names derived from co-occurrence networks in
\wikipedia.

In Section \ref{sec:internetwork}, co-occurrence networks derived from
\wikipedia in different languages are compared. The eigenvector
centrality based comparative analysis revealed language specific
features, both for given names and city names. This result suggests
that features derived from co-occurrence networks of different
languages of \wikipedia can be used to train classifiers for detecting
language specific entities. We plan to implement and evaluate such
classifiers for labeling given names according to their language
association and incorporate the obtained results in the \nameling.

Section~\ref{sec:task} focused on the evaluation of different
similarity metrics, relative to a respectively fixed notion of
semantic relatedness. Firstly, the considered list of similarity
functions is not exhaustive. Especially, all considered similarity
functions only considered local features, \ie based on the direct
neighborhood. Accordingly, we will evaluate more similarity
functions. But also for the reference relation more alternatives have
to be considered. From a practical point of view, different types of
relatedness among given names are of interest, as, \eg language
specific variants or originating cultural background. Different
similarity functions may capture different forms of semantic
relatedness. Furthermore, the experimental set up in
Section~\ref{sec:task} can be used to formulate a machine learning
task, aiming at optimizing a similarity function based on features
derived from the co-occurrence networks.

We will apply the \nameling's usage statistics for evaluating
different similarity functions with respect to human interaction in a
specific recommender scenario.

%
%
\bibliographystyle{abbrv}\enlargethispage*{2\baselineskip}
\bibliography{bibliography}

\end{document}